\documentclass[useAMS,usenatbib]{mn2e}

\usepackage{amssymb}
\usepackage{amsmath}
\usepackage{epsfig}
\usepackage{multicol}

\usepackage{graphicx}
\usepackage{dcolumn}
\usepackage{bm}
\usepackage{epstopdf}

\newcommand{\mBD}{$\mu$BD}
\newcommand{\mBDs}{$\mu$BDs}

\newcommand{\half}{{\frac{1}{2}}}
\newcommand{\Hii}{H{\small II}}
\newcommand{\dHe}{$^3$He}

\newcommand{\myskip}[1]{}

\renewcommand{\d}{{\rm d}}

\newcommand{\BEQ}{\begin{eqnarray}}
\newcommand{\EEQ}{\end{eqnarray}}
\newcommand{\BEA}{\begin{eqnarray}}
\newcommand{\EEA}{\end{eqnarray}}

\date{Accepted ... Received ...; in original form ...}

\title{Explanation of the Helium-3 problem}

\author[T.M. Nieuwenhuizen]
{Theo M. Nieuwenhuizen$^{1,2}$\thanks{E-mail: t.m.nieuwenhuizen@uva.nl} \\
$^{1}$ Center for Cosmology and Particle Physics, New York University, 
 4 Washington Place, New York, NY 10003, USA \\
$^{2}$Institute for Theoretical Physics, University of Amsterdam,
Science Park 904, P.O. Box 94485, 
1090 GL  Amsterdam, The Netherlands}

\begin{document}
\maketitle

\begin{abstract}
One of the tests of nucleosynthesis theory is the $^3$He abundance in the Galaxy. 
$^3$He$^+$ is observed through its 3.46 cm hyperfine level in \Hii \  regions and 
the $^3$He/H ratio  compares well with theory.
Since $^3$He \ can be created or destroyed in nuclear reactions, one would expect that its abundance shows a trend with
the amount of such reactions, so with distance to the Center of the Galaxy and with metallicity. Such trends are lacking in observations.
This is explained by assuming that the \Hii \  clouds are recently formed out of the primordial micro brown dwarfs of earth mass
predicted by gravitational hydrodynamics. If indeed existing, they would preserve their primordial $^3$He/H  ratio and spread
this when evaporating into \Hii \ clouds, independent of the location in the Galaxy. 

In the development of the argument, it is also explained that wide binaries do not rule out the MACHO dark matter 
predicted by gravitational hydrodynamics, but are rather immersed as visible partners in Jeans clusters of dark micro brown dwarfs.
\end{abstract}


\section{Introduction}

Nucleosynthesis in the WMAP era is said to be a parameter-free theory, see e. g. \citet{Steigman2007}. 
With cosmology described by the Friedman-Lemaitre-Robertson-Walker metric and the number of neutrino species known, 
accurate predictions  can be made, ready for comparison with observations. We will discuss one of these tests, 
the one for \dHe,  that is in agreement with theory, but has a puzzling aspect.

\section{The helium-3 problem}

$^3$He is produced in nucleosynthesis and the observed 
abundance [\dHe/H]$\approx1.0\cdot 10^{-5}$ agrees with theory. But not all is well, so let us explain the way to observe it.
The $^3$He nucleus consists of two protons and a neutron, so it has  spin $\frac{3}{2}$ or $\half$.
The singly ionized $^3$He$^+$ contains one electron, thus the singlet spin-$\half$ state brings an analogy with the neutral H-atom.
Indeed, there is an analog of the 21-cm spin-flip transition, which occurs roughly at 21/$Z^2$=21/4 cm,
more precisely at 3.46 cm. The emission by this transition line provides an observational signature of singly ionized
$^3$He located in regions of ionized H gas (so-called \Hii\ regions) and in planetary nebulae that also contain ionized H and $^3$He.

 Hii \ regions have been created rather recently in the cosmic history (\citealt{Rood2002, Bania2007}), 
and are thus expected to reflect the chemical history of the surroundings. $^3$He can be created or destroyed in nuclear reactions. 
One would thus expect its local concentration
in the Galaxy to depend on the amount of reactions that have taken place, a measure for which is the local metallicity.
Since near the Center more reactions took place, there is a clear gradient in the metallicity, so one also 
expects a significantly higher $^3$He concentration near the Center than in the outskirts. 
But though observations show fluctuating concentrations, there is no statistical trend with distance or with
metallicity. Specifically, low values of [\dHe/H]$\approx1.0\cdot 10^{-5}$ are observed at 4, 9, 10, 12 and 16
kpc from the Center, while up to two-three times higher values are observed in between (\citealt{Rood2002}).

The important point for the present paper is that the \Hii \  regions show no evidence for stellar \dHe \ enrichment during the last 4.5 Gyr;
the low values thus being found at any distance from the Center is called {\it  the} \dHe \ {\it problem} (Bania et al. 2007).
This situation creates a paradox, since it calls for a delicate balance
between creation and destruction {\it independent of the amount of reactions} (the metallicity).
To be compatible with this result, Galactic chemical evolution models (\citealt{Tosi2000}) require that
$\sim90\%$ of solar analog stars are non-producers of \dHe. But the two best observed planetary nebulae 
do have higher concentration (\citealt{Rood2002}), and consequently should then both be members of the 10\% class
of stars that do produce \dHe \ -- which seems unlikely.
For a review of the current status of \dHe  \ evolution, see \citet{Romano2003}.

\section{Micro brown dwarfs from Gravitational hydrodynamics}

Gravitational hydrodynamics is an approach that stresses the role of turbulence in cases of small viscosity
(more precisely, large Reynolds number) and the possibility of viscous structure formation of turbulent flows
on scales where the Reynolds number is low. It is well understood that at the transition from plasma to neutral gas
(decoupling or recombination), the newly formed gas breaks up at the Jeans scale,
forming Jeans clumps of some 600,000 $M_\odot$. It was put forward by \citet{Gibson1996} that the Jeans clumps themselves 
fragment at the viscous scale into objects of earth mass, turning the Jeans gas clumps into Jeans clusters of some $2\cdot10^{11}$
micro brown dwarfs  (\mBDs) of earth mass. The Galaxy is predicted to contain about two million Jeans clusters in its halo that make 
up the full Galactic dark matter (\citealt{Gibson1996}, \citealt{NSG2010}).

These objects are belong to the class called MACHOs (Massive Astrophysical Compact Halo Objects). 
They have been detected in quasar micro lensing (\citealt{Schild1996})  and the observed signature has 
approximately equal positive and negative events (\citealt{Schild1999}),
 uniquely indicating microlensing by a population at unit optical depth.
However, in direct searches in front of the Magellanic clouds they have not been detected.  Inspection of the literature
in this field reveals  that this low mass scale was covered only in one paper (\citealt{Renault1998}).
This reports observations carried out in the early and mid nineties using a telescope of 40 cm in diameter.
Later MACHO searches did not cover this mass range, so the controversy between the Schild detection and the
Renault et al. non-detection was never resolved. It is planned to redo the MACHO search in front of the Magellanic clouds
using a much larger telescope (Schild 2011). They will be searched in what is normally called cirrus dust clouds,
but what could just be an agglomeration of dark Jeans clusters. One further indication for this bold assumption is that the
temperature of ``cirrus dust'' is about 15 K (\citealt{Veneziani2010}), i. .e., near the H triple point of 13.8 K. 
While cold dust theories have no explanation why the dust should minimally have  this temperature,
the required release of latent heat of these compact hydrogen clouds  would then keep them long at this temperature,
before finally freezing.

Analysis of wide binaries in the Galaxy halo has ruled out any MACHO dark matter component more heavy than 43 $M_\odot$ 
(Yoo et al. 2004). That this conclusion is somewhat too strong  (\citealt{Longhi2010} ), will be of little help in our situation.
For  a typical wide separation of the components,  the major axis is of the order $a=$ 18,000 AU $=0.09$ pc.
In our case the relevant MACHO \ objects for this application are  the individual Jeans clusters of 600,000 $M_\odot$ 
and radius of 1.4 pc,  so they would completely disrupt the binaries -- if they were on their own.
However, we now point out that these halo wide binaries - being much smaller than the Jeans clusters - 
must lie inside Jeans clusters themselves, as a result of which they must be much more stable than commonly expected. 
So we predict that these wide binaries have a center of mass motion of ca 200 km/s, a mutual speed of ca 20 km/s and are 
embedded in a baryonic dark matter matrix of ca 600,000 $M_\odot$. These features  can be tested.
Clearly, stabilized inside Jeans clusters, wide binaries do not rule out the MACHO dark matter.

One would expect that the fragmentation of the Jeans cloud has been seen in numerics. When Truelove et al. (1997) 
noticed instabilities in their simulations, they dismissed them as unphysical and built in Jeans filters to remove them.
Recently instabilities at the $10^{-3}-10^{-4}M_\odot$ level were observed in the simulation of the first stars
in turbulent gas clouds (Clark et al. 2011): the gas first fragments and then the fragments aggregate to form the stars. 
The authors carefully explain how the gas can cool, namely grace to the formation of molecular hydrogen (H$_2$) with help of a 
small fraction of free electrons still present, so that heat can be radiated away through H$_2$ levels. This allows the gas to cool
locally after which it can fragment. Though the authors did not carefully sort out what would be the minimal fragmentation scale,
 their fragmentation at  the  $10^{-3}-10^{-4}M_\odot$ level is a already strong support for the gravitational hydrodynamics picture 
 and indeed points towards its predicted earth mass MACHOs.
 
Nuclear synthesis attributes enough matter for the \mBDs. About 4.5\% of the critical
density of the Universe is baryonic. At best 0.5 \% is luminous, and some 2--3\% is observed in X-ray gas. 
The {\it missing baryon problem} refers to the fact that most of the baryons are unaccounted for, about 60\% is missing at cosmic scales; 
an inventory of baryons in and around the Milky Way reveals at best 25\% of the expected baryons.
Though they are believed to be located in unobserved relatively cool X-ray clouds, this need not be the full or the only explanation.
Indeed, the missing dark baryons may  be locked up in \mBDs. 
Radiation from what is commonly called ``cold cirrus clouds'' or ``cold cirrus dust'' is observed to have a temperature of 15 K (Veneziani et al. 2010),
or, more generally, temperatures between 40 K and 15 K  (\citealt{Amblard2010}).  Dust models have no explanation for the minimum of 15 K.
But the radiation  may actually arise from the thermal atmospheres of \mBDs \ (\citealt{NSG2010}).
The fact that H has a critical point at 33 K and a triple point at 13.8 K coincides with many observations of ``cold dust'' temperatures
by the Herschel telescope between 15 and 40 K, with the lower values condensing near 15 K, see Figures 2 and 3 of  \citet{Amblard2010}. 

The picture of baryonic dark matter locked up in \mBDs\ grouped in -- mostly -- dark Jeans clusters (JCs) explains another, not often mentioned
problem: {\it the missing Jeans clusters problem}. Indeed, it is agreed that after the decoupling all gas fragments in Jeans clumps,
but where are they now? Gravitational hydrodynamics asserts that they just constitute the halo dark matter of galaxies.

From the point of view of galactic structures there is a lot of support for the picture of Jeans clouds consisting of micro brown dwarfs.
Galactic rotation curves flatten when the ca $2\cdot 10^6$ JCs of the Galaxy have an isothermal distribution $\rho(r)\approx v^2/2\pi G r^2$,
where $v$ is the velocity dispersion, about 200 km/s for Jeans clusters in the Galaxy and 20 km/s for \mBDs \ inside a Jeans cluster.
The Tully-Fisher and and Faber-Jackson relations follow if one assumes that star formation arises when JCs heat each others
mBDs \ by tidal forces when they come within a certain radius of each other, which brings the proportionality
{\it Luminosity} $\sim \#\, {\rm stars}\, \sim \int\d^3r\,\rho^2(r)\sim v^4/R_\ast$ for some characteristic radius $R_\ast$
 ~(\citealt{NGS2009}).

In galaxy merging the observed young globular clusters may not or not only appear due to tidal disruption, and its unknown star formation 
process, but also due to tidal heating of the mBDs, which expand and can form stars millions of years after the merging process has taken place
 (\citealt{NSG2010}). Thus the galaxy merging process can transform, along the merging path, dark Jeans clusters in situ in 
 the observed bright young globular clusters.

 Mysterious  radio events were reported by Ofek et al. (2010). They are frequent (1000/square degree/year), radio loud ($>$ 1Jy),
 and have neither a precursor nor a follow up, and have no detectable counterparts in the infrared, visible or X-ray spectrum. 
 Within isothermal modeling,  they have been connected to merging of \mBDs  \ inside Jeans clusters, 
 and the event duration of more than half an  hour to several days
 allowed to estimate their radius as 3 $\times$ (the duration in days) $\times$ (the solar radius) (\citealt{NSG2010}).

The theory of star formation that has been inconclusive for long, strongly benefits from the principle that star (and planet) formation
arises from aggregation of \mBDs \ (Gibson and Schild 2011). The {\it iron planet core problem} acknowledges that
iron planet cores, like in the earth, are difficult to explain, since iron is mostly observed in oxides. But \mBDs \ with their size larger than the Sun
and weighing only as much as the Earth, can collect the intergalactic iron dust as ``vacuum cleaners'', while the H atmosphere is all too eager 
to dissolve the oxides for making water, after which the iron can sink to the center. Iron cores are then explained from glueing small iron cores
in the aggregation of the \mBDs \  to form planets  (\citealt{NSG2010}).

 
 A relation between globular clusters and black holes was discussed (\citealt{NSG2010}) and will be expanded elsewhere, 
 in particular in connection with role of Jeans clusters in solving the so-called last parsec problem in black hole merging processes.
 
Nowadays galaxies are observed at redshifts basically up to $z=10$. This being considered before 
the reionization era, they should not be visible but immersed in H clouds, so that many unobserved weak galaxies are invoked, 
which should ionize the gas and create a ``pencil of visibility'' in our direction. 
The galaxy UDFy-38135539 was established to have redshift $z=8.55$, so we see light that it emitted 600 million years 
after the big bang~(\citealt{Lehnert2010}).
The authors say in the abstract that a significant contribution from other, probably fainter galaxies nearby, is needed
to explain its visibility. This is a deus ex machina.
\mBDs, on the other hand, offer a more obvious explanation: the hydrogen is locked up
in condensed objects, so most of the space is empty and transparent  (\citealt{NSG2010}).

One observes that the amount of dust and the rate of star formation had a peak at redshift $z\sim2$ and are smaller in recent times.
This higher amount of dust content in massive galaxies at higher redshift 
is difficult to explain in standard dust evolution models (\citealt{Dunne2010}).
But it  finds an easy explanation if one accepts that the ``dust'' radiation is produced by the atmospheres of \mBDs.
In the course of time, they are more and more  used up because they coagulate to form heavier objects and stars.
Consequently, less fuel for star formation is  ultimately available and less \mBD \ outer surface exists to emit ``dust'' radiation.

\section{The answer to the Helium-3 problem}

Within gravitational hydrodynamics it is natural to identify 
the \Hii \  clouds with partly or fully evaporated Jeans clusters, in which a large fraction of the \mBDs, if not all, evaporated 
by heating in a strong star forming region. In such a case the \mBDs \  of a  Jeans cluster can expand into a gas cloud.
Their would have kept the primordial $^3$He content up to then, and it would not change when they evaporated into \Hii \ clouds, 
so the primordial value would appear independent of their
location in the Galaxy and independent of the local metallicity, which explains the \dHe \  problem.


 The most massive and largest \Hii \ region in the Local Group is  the region 30 Doradus in the Large Magellanic Cloud. 
 It has been carefully studied recently (Lopez et al 2008).
In addition to a large ionized gas mass $\sim 8 \cdot 10^5M_\odot$ (Kennicutt 1984), 
the 30 Doradus nebula also has $\sim 10^6M_\odot$ of CO (Johansson et al. 1998).
This \Hii \ mass is approximately the mass of a single Jeans cluster, supporting the just mentioned 
gravitational hydrodynamics picture, while the CO content exhibits a large metallicity.

In conclusion, in gravitational hydrodynamics many problems, like the 15 K cold dust temperature, the visibility
of early galaxies and the \dHe \ problem, 
find a simple explanation.


\subsection*{Acknowledgment}
It is a pleasure to thank David Hogg for pointing us at the wide binary problem.

\newcommand{\ital}[1]{}


\begin{thebibliography}{99}
\small 


\bibitem[\protect\citeauthoryear{Amblard et al.}{2010}]{Amblard2010}
\ital{arXiv:1005.2412 Herschel-ATLAS: Dust temperature and redshift distribution of SPIRE and PACS detected
 sources using submillimetre colours}
 A. Amblard, A. Cooray, P. Serra, P. Temi, E. Barton, M. Negrello, R. Auld, M. Baes, I.K. Baldry, S. Bamford, A. Blain, J. Bock, 
 D. Bonfield, D. Burgarella, S. Buttiglione, E. Cameron, A. Cava, D. Clements, S. Croom, A. Dariush, G. de Zotti, S. Driver, J. Dunlop,
  L. Dunne, S. Dye, S. Eales, D. Frayer, J. Fritz, Jonathan P. Gardner, J. Gonzalez-Nuevo, D. Herranz, D. Hill, A. Hopkins, D. H. Hughes,
   E. Ibar, R.J. Ivison, M. Jarvis, D.H. Jones, L. Kelvin, G. Lagache, L. Leeuw, J. Liske, M. Lopez-Caniego, J. Loveday, S. Maddox, 
   M. Michalowski, P. Norberg, H. Parkinson, J.A. Peacock, C. Pearson, E. Pascale, M. Pohlen, C. Popescu, M. Prescott, A. Robotham, 
   E. Rigby, G. Rodighiero, S. Samui, A. Sansom, D. Scott, S. Serjeant, R. Sharp, B. Sibthorpe, D.J.B. Smith, M.A. Thompson, 
   R. Tuffs, I. Valtchanov,  E. Van Kampen, P. Van der Werf, A. Verma, J. Vieira, and C. Vlahakis, 2010, A\&A 518, L9


\bibitem[\protect\citeauthoryear{Bania et al.}{2007}]{Bania2007}
Bania, T. M., Rood, R. T.,  Balser, D. S., 2007,
\ital{The Milky Way 3-Helium Abundance, }
Space Science Reviews, 130, 53-62.

\bibitem[\protect\citeauthoryear{Clark et al.}{2011}]{Clark2011}
Clark, P. C.,  Glover, S. O. C., Klessen, R. S.,   Bromm, V., 2011, ApJ 727, 110



\bibitem[\protect\citeauthoryear{Dunne et al.}{ 2010}]{Dunne2010}
L. Dunne, H. Gomez, E. da Cunha, S. Charlot, S. Dye, S. Eales, S. Maddox, K. Rowlands, D. Smith, R. Auld, M. Baes, 
D. Bonfield, N. Bourne, S. Buttiglione, A. Cava, D. Clements, K. Coppin, A. Cooray, A. Dariush, G. de Zotti, S. Driver, 
J. Fritz, J. Geach, R. Hopwood, E. Ibar, R. Ivison, M. Jarvis, L. Kelvin, E. Pascale, M. Pohlen, C. Popescu, E. Rigby, 
A. Robotham, G. Rhodighiero, A. Sansom, S. Serjeant, P. Temi, M. Thompson, R. Tuffs, P. van der Werf, C. Vlahakis, 
2010,  arXiv:1012.5186
\ital{Herschel-ATLAS: Rapid evolution of dust in galaxies in the last 5 billion years}



\bibitem[\protect\citeauthoryear{Gibson}{1996}]{Gibson1996}
Gibson, C. H. 
\ital{Turbulence in the Ocean, Atmosphere, Galaxy, and Universe.}
1996 {Appl. Mech. Rev.} {\bf 49}, 299--315 


\bibitem[\protect\citeauthoryear{Johansson et al.}{1998}]{Johansson1998}
\ital{Results of the SEST key programme: CO in the magellanic clouds - VII. 30 Doradus and its southern HII regions}
Johansson, L. E. B. , Greve A., Booth, R. S., Boulanger, F., Garay, G., de Graauw, T., Israel, F. P., Kutner, M. L., Lequeux, J., 
Murphy, D. C., Nyman, L. A., Rubio, M., 1998, A\&A, 331, 857-872
\ital{Abstract: We have mapped the (CO)-C-12(1-0) emission from the 30 Doradus region in the Large Magellanic Cloud with the 
Swedish-ESO Submillimetre Telescope (SEST). The regions investigated include the central part of the 30 Dor nebula, and the 
southern H II regions N 158C, N 159, and N 160. In addition, a few prominent CO clouds have been studied in the (2-1) and (3-2) 
transitions of CO. The southern area shows the strongest (CO)-C-12(1-0) emission, a factor of 3 higher than towards the central part of 30 Dor. 
A non-LTE analysis of the CO emission from a sample of clouds indicate kinetic temperatures between 10 and 50 K; the highest temperatures 
are found close to the 30 Dor nebula. We have estimated the CO-H-2 conversion factor for the two areas, separately, under the assumption 
that the virial mass, determined from CO parameters, reflects the total molecular mass. We find an unexpectedly small difference between 
the two areas. This is explained as a bias effect.}


\bibitem[\protect\citeauthoryear{Kennicutt}{1996}]{Kennicutt1984}
Kennicutt, Jr., R. C. 1984, ApJ, 287, 116

\bibitem[\protect\citeauthoryear{Lehnert et al.}{ 2010}]{Lehnert2010}
Lehnert , M. D., Nesvadba, N. P. H., Cuby, J. -G., Swinbank, A. M., Morris, S., Clement, B.,  Evans, C. J., Bremer, M. N., Basa, S.,
\ital{Spectroscopic confirmation of a galaxy at redshift z = 8.6.}
2010, {Nature} {\bf 467}, 940-942

\bibitem[\protect\citeauthoryear{Longhitano \& Binggeli}{ 2010}]{Longhi2010} 
\ital{The stellar correlation function from SDSS - A statistical search for wide binary stars}
Longhitano, M.,  Binggeli, B. , 2010, A\&A, 509, A46 



\bibitem[\protect\citeauthoryear{Lopez et al.}{2008}]{Lopez2008}
L. A. Lopez, M. R. Krumholz, A. D. Bolatto, J. Xavier Prochaska, E. Ramirez-Ruiz, 2008,
\ital{What Drives the Expansion of Giant HII Regions?: A Study of Stellar Feedback in 30 Doradus,}
arXiv:1008.2383.

\bibitem[\protect\citeauthoryear{Nieuwenhuizen et al.  }{2009}]{NGS2009}
Nieuwenhuizen, T. M., Gibson, C. H., Schild, R. E., 
\ital{Gravitational hydrodynamics of large-scale structure formation.}
2009, { Europhys. Lett.},  {\bf 88}, 49001,1--6


\bibitem[\protect\citeauthoryear{Nieuwenhuizen et al. }{2010}]{NSG2010}
Nieuwenhuizen, T. M.,  Schild, R. E., Gibson, C. H., 
\ital{Gravitational hydrodynamics of large-scale structure formation.}
2009, { Europhys. Lett.}  {\bf 88}, 49001,1--6


\bibitem[\protect\citeauthoryear{Ofek et al.}{ 2010}]{Ofek2010} 
Ofek, E. O. et al. \  
\ital{Long-duration radio transients lacking optical counterparts are possibly galactic neutron stars.}  
2010, ApJ, \ {\bf 711},  517--531 



\bibitem[\protect\citeauthoryear{Renault et al.}{1998}]{Renault1998} 
Renault, C. et al. \ 
\ital{Search for planetary mass objects in the Galactic halo through microlensing.}
1998 A\&A \ {\bf 329} 522--537

\bibitem[\protect\citeauthoryear{Romano et al.}{2003}]{Romano2003}
Romano D, Tosi M, Matteucci F, Chiappini C. MNRAS 346:295 (2003).


\bibitem[\protect\citeauthoryear{Rood et al.}{2002}]{Rood2002}
Rood, R. T.; Bania, T. M., Balser, D.S., 2002,
\ital{ The Saga of 3He}
Science, 295, 804-805 


\bibitem[\protect\citeauthoryear{Schild}{1996}]{Schild1996}
Schild, R. E.,
\ital{Microlensing variability of the gravitationally lensed quasar Q0957+561 A,B.}
1996, ApJ, \ {\bf 464}, 125--130

\bibitem[\protect\citeauthoryear{Schild}{1999}]{Schild1999}
Schild, R. E., 
\ital{A wavelet exploration of the Q0957+561 A, B brightness record. }
1999, ApJ 514, 598Ð606 

\bibitem[\protect\citeauthoryear{Schild}{2011}]{Schild2011}
Schild, R. E.,  2011, private communication.


\bibitem[\protect\citeauthoryear{Steigman}{2007}]{Steigman2007} 
Steigman, G., 2007,
\ital{Primordial Nucleosynthesis in the Precision Cosmology Era,}
Annual Review of Nuclear and Particle Science 57: 463-491.

\bibitem[\protect\citeauthoryear{Truelove et al.}{1997}]{Truelove1997}  
Truelove, J. K., Klein, R. I.,  McKee, C. F., Holliman,  J. H., Howell, L. H., Greenough, J. A.,
\ital{The Jeans condition. A new constraint on spatial resolution in simulations 
of isothermal self-gravitational hydrodynamics. }
1997, ApJ, 489, L179







\bibitem[\protect\citeauthoryear{Tosi}{2000}]{Tosi2000}
M. Tosi, in The Light Elements and Their Evolution, Proceedings of IAU Symposium 198, ed. by L. da Silva,
M. Spite, J.R. de Medeiros (ASP, San Francisco, 2000), pp. 525-532



\bibitem[\protect\citeauthoryear{Veneziani et al.}{ 2010}]{Veneziani2010} 
Veneziani, M.,
Ade, P. A. R.,
Bock, J. J.,
Boscaleri, A.,
Crill, B. P.,
de Bernardis, P.,
De Gasperis, G.,
de Oliveira-Costa, A.,
De Troia, G.,
Di Stefano, G.,
Ganga, K. M.,
Jones, W. C.,
Kisner, T. S.,
Lange, A. E.,
MacTavish, C. J.,
Masi, S.,
Mauskopf, P. D.,
Montroy, T. E.,
Natoli, P.,
Netterfield, C. B.,
Pascale, E.,
Piacentini, F.,
Pietrobon, D.,
Polenta, G.,
Ricciardi, S.,
Romeo, G.,
Ruhl, J. E.,
\ital{Properties of galactic cirrus clouds observed by BOOMERanG. }
2010, ApJ, { 713}, 959--969 

 



\bibitem[\protect\citeauthoryear{Yoo et al.}{2004}]{WideBinaries}
Yoo, J., Chanam\'e, J. and Gould, A. 2004.
\ital{The end of the MACHO era: limits on halo dark matter from stellar halo wide binaries,}
Astrophysical Journal {\bf 601}, 311-318.



\end{thebibliography}
\end{document}